\documentclass[lettersize,journal]{IEEEtran}

\usepackage{balance}
\usepackage[utf8]{inputenc}
\usepackage{graphicx}
\usepackage{ulem}
\usepackage{graphicx}
\usepackage{float}
\usepackage{array}
\usepackage{tabularx}
\usepackage{multirow}
\usepackage{amsmath}
\usepackage{amssymb}
\usepackage{threeparttable}
\usepackage{setspace}
\usepackage{color}
\usepackage{colortbl}
\usepackage{xcolor}
\usepackage{soul}
\usepackage{booktabs}
\usepackage{makecell}
\usepackage{algorithm} 
\usepackage{algpseudocode}
\usepackage{hyperref}
\usepackage{caption}
\usepackage{enumitem}
\usepackage{listings}
\usepackage{xcolor}
\usepackage{picins}
\usepackage{verbatim}
\usepackage{tikz}
\usepackage{xcolor}

\def\BibTeX{{\rm B\kern-.05em{\sc i\kern-.025em b}\kern-.08em
    T\kern-.1667em\lower.7ex\hbox{E}\kern-.125emX}}
\usepackage[utf8]{inputenc}
\usepackage[framemethod=TikZ]{mdframed}
\usepackage{tcolorbox}

\newcommand*\circled[1]{\tikz[baseline=(char.base)]{
            \node[shape=circle,fill,inner sep=1pt] (char) {\textcolor{white}{#1}};}}

\author{Dianshu~Liao,
        Shidong~Pan,
        Xiaoyu~Sun,
        Xiaoxue~Ren,
        Qing~Huang,
        Zhenchang~Xing,
        Huan~Jin,
        Qinying~Li
\IEEEcompsocitemizethanks{
\IEEEcompsocthanksitem D. Liao, X. Sun are with School of Computing, Australian National University, Australia.
\IEEEcompsocthanksitem Q. Huang, the corresponding author, is with the School of Computer Information Engineering at Jiangxi Normal University, China (e-mail: qh@jxnu.edu.cn).
\IEEEcompsocthanksitem S. Pan and Z. Xing are with CSIRO's Data61, Australia, and the School of Computing, Australian National University, Australia.
\IEEEcompsocthanksitem X. Ren is with the State Key Laboratory of Blockchain and Data Security, Zhejiang University. Also with Hangzhou High-Tech Zone (Binjiang) Institute of Blockchain and Data Security, China. 
\IEEEcompsocthanksitem H. Jin, Q. Li are with the Jiangxi University of Technology, China.
}
}

\begin{document}

\title{$\mathbf{A^3}$-CodGen : A Repository-Level Code Generation Framework for Code Reuse with Local-\underline{A}ware, Global-\underline{A}ware, and Third-Party-Library-\underline{A}ware}

% make the title area
\maketitle

\begin{abstract}
LLM-based code generation tools are essential to help developers in the software development process. Existing tools often disconnect with the working context, i.e., the code repository, causing the generated code to be not similar to human developers. 
In this paper, we propose a novel code generation framework, dubbed 
\textbf{$A^3$}-CodGen, to harness information within the code repository to generate code with fewer potential logical errors, code redundancy, and library-induced compatibility issues.
We identify three types of representative information for the code repository: local-aware information from the current code file, global-aware information from other code files, and third-party-library information. 
Results demonstrate that by adopting the \textbf{$A^3$}-CodGen framework, we successfully extract, fuse, and feed code repository information into the LLM, generating more accurate, efficient, and highly reusable code. 
The effectiveness of our framework is further underscored by generating code with a higher reuse rate, compared to human developers.
This research contributes significantly to the field of code generation, providing developers with a more powerful tool to address the evolving demands in software development in practice.
\end{abstract}


\section{Introduction}

\IEEEPARstart{I}n the realm of Software Engineering, the advent of Large Language Models (LLMs) has revolutionized various tasks, including code generation~\cite{svyatkovskiy2020intellicode,li2022competition,puschel2005spiral,chen2022codet,sun2020treegen}.
The automatic generation of code by these models significantly reduces development time and effort~\cite{garcia2008association}.
Specifically, in the process of evolving software projects, software engineering developers are frequently tasked to add new features and functionality~\cite{cusumano1999software}.
LLMs such as ChatGPT~\cite{ChatGPT}, GitHub Copilot~\cite{GithubCopilot}, Codex~\cite{chen2021evaluating} are capable of automatically generating code based on the developers' requirements (e.g., natural language descriptions of functions and function definitions).

A significant gap still persists between the code generated by LLMs and that crafted by human developers.
When adding new functionality to a specific module, developers already had sufficient knowledge about the current task and the entire code repository.
Also, according to a previous study~\cite{ciniselli2023source}, developers often expect that code generation tools could be aware of more context/knowledge by using specific frameworks/libraries~\cite{liucodegen4libs}. 
In addition, they hope the generated code maximum utilizes specific third-party libraries (e.g., invocation of an API in a library)~\cite{zan2022cert,li2021honeysuckle,zan2023private}.
Above all, in contrast to straightforward LLM-based code generation, developers especially consider three primary categories of information:

\circled{1} \textbf{Local Information}:
This category encompasses information within the current working context, including defined function signatures and variables.
It also covers the relative file paths of the module within the project, such as the module's Fully Qualified Names (FQNs), like \textit{unstructured.documents.html}. 
Local information is crucial to ensuring the generated code utilizes accurate variables and operates harmoniously with the other functions in the project.

\circled{2} \textbf{Global Information}:
This category involves functions from other code files within the same repository, accessible via fully qualified import API names.
A previous research highlighted that only around 30\% of methods operate independently in open-source projects~\cite{yu2023codereval}.
Hence, these external functions are essential for implementing new functions without duplicating code.
For example, developers import functions ``from \textit{unstructured.partition.text\_type} import \textit{is\_bulleted\_text}'' from other modules, instead of writing a function to judge ``bulleted text'' from scratch.

\circled{3} \textbf{Third-Party-Library Information}: This category includes third-party libraries pre-installed in the code repository's environment. 
These libraries can be accessed directly through import statements, enhancing functionality without extra dependencies.

Unfortunately, none of the existing code generation methods take the above three types of information into consideration simultaneously.
As the most common code generation tool, ChatGPT~\cite{ChatGPT} demonstrates commendable coding ability~\cite{zhuang2023efficiently} but relies solely on user input. 
GitHub Copilot~\cite{GithubCopilot}, which explicitly utilizes local information spanning the current working context, often struggles to harness the full potential of global information presented in the entire code repository.\footnote{\href{https://githubnext.com/projects/copilot-view/}{https://githubnext.com/projects/copilot-view/}}
When these methods disregard local and global module information, they risk omitting crucial logic during code generation~\cite{gifany2013logic}. 
For instance, they might forget to extract text from a tag before attempting to identify it as a bullet point (\textit{is\_bulleted\_text}).
Furthermore, neglecting third-party library information may cause the LLMs to randomly select third-party libraries and generate code based on them.
If the selected libraries third-party are unsupported under the given environment, it can lead to compatibility issues, such as ``\textit{ModuleNotFoundError: No module named xxx.}''

Therefore, to encounter the insufficient information awareness issue in LLM code generation tasks, we introduce novel \textbf{$\mathbf{A^3}$-CodGen} (Local \textbf{\underline A}ware, Global \textbf{\underline A}ware, and Third-Party-Library \textbf{\underline A}ware \textbf{\underline{Cod}}e \textbf{\underline {Gen}}eration).
This framework systematically mines and utilizes external knowledge from the local module, global modules, and third-party libraries.
First, our framework begins by acquiring foundational information about functions and third-party libraries within the code repository. This initial step provides a solid function base and a third-party-library base that supports subsequent information retrieval processes.
Next, it conducts knowledge retrieval from the function base, the third-party library base, and the current file. 
These retrieved insights are systematically organized into three distinct categories of code repository-aware knowledge.
Subsequently, the framework seamlessly fuses these three types of code repository-aware knowledge into a cohesive prompt.
This prompt serves as the input to the foundational LLM, enabling the model to generate code that closely resembles the work of human developers.

Our experiments are driven by two primary objectives:
1) To investigate the effectiveness of three distinct types of code repository information in assisting the model with code generation for specific code repositories.
2) To assess whether our framework can efficiently extract and utilize these three types of information.
We crawl and filter 29 representative Python code repositories from PyPI, assembling 13,784 functions.
To ensure a diverse benchmark dataset, we conduct a sampling process, selecting representative functions from each repository to total 383 functions.
We subsequently label these benchmarks and the model outputs generated under different configurations to evaluate the performance of $A^3$-CodGen.

Our experimental results confirm the efficacy of incorporating three types of code repository-awareness knowledge in aiding the foundational model to reuse code repository knowledge. 
The $A^3$-CodeGen framework excels in recognizing and utilizing these three types of code repository knowledge, leading to exceptional performance metrics in terms of local reuse, global reuse, and third-party library reuse.
Thanks to its outstanding code reuse capabilities, our framework is capable of generating concise code that seamlessly integrates with the code repository's working environment. 
Furthermore, we identify optimal configurations for local and global information, enabling the model to effectively leverage essential knowledge while avoiding the potential pitfalls of information overload.
By comparing $A^3$-CodGen with GitHub CoPilot, we find that while both perform similarly in local reuse, $A^3$-CodGen surpasses CoPilot in global reuse and third-party library reuse.
We propose potential methods for improving industry products like CoPilot by incorporating our approach.

The main contributions of this paper are as follows.

\begin{itemize} [leftmargin=*]
    \item To the best of our knowledge, we are the first to conduct an extensive empirical study, systematically investigating the usefulness of local information, global information, and third-party-library information in code generation tasks. 
    
    \item We introduce a novel framework, dubbed \textbf{$\mathbf{A^3}$-CodGen Framework}, effectively knowledge mining and knowledge utilization, aiding models in better understanding and fully utilizing the knowledge within code repositories.

    \item We construct a new benchmark dataset, namely \textbf{RepoEval}, to reflect the effectiveness on code-repository-aware code generation. Results demonstrate the validity of the three types of code repository information and show that our framework can effectively mine and utilize this knowledge.

    \item We open-source our replication package~\cite{our_repo}, including the dataset, source code, and experimental results, for follow-up research.
\end{itemize}
\section{Motivation}

\begin{figure*}%[H]
    \centering
    \includegraphics[width=1\textwidth]{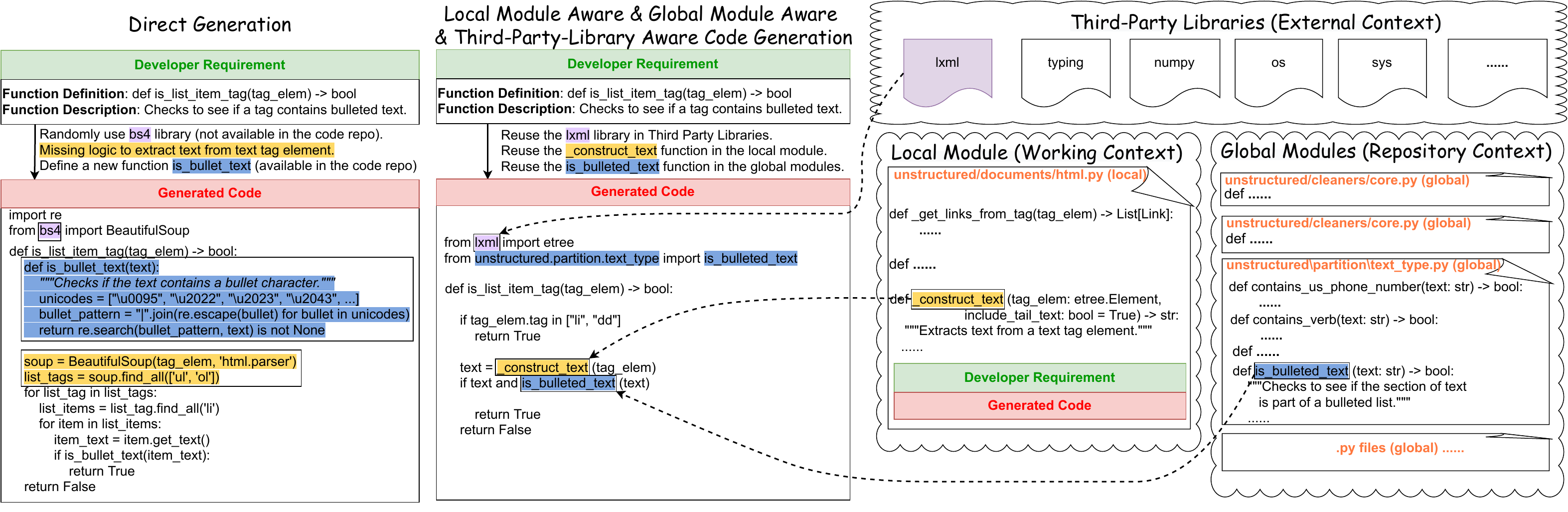}
    \caption{Knowledge-Starved Code Generation vs. Code Repository Aware Code Generation.}

    \label{fig:killing example}
    % \vspace{-10pt}
\end{figure*}

\subsection{Limitation of Existing Code Generation Approaches}
\label{sec_motivation_limitation}
Benefiting from the rapid development of Transformer structures and pre-trained LLMs in the early 2020s, code-LLMs experienced significant growth. Transformer code models, exemplified by milestone works such as CodeT5~\cite{wang2021codet5}, CodeBERT~\cite{feng2020codebert}, and GraphCodeBERT~\cite{guo2020graphcodebert}, dominated the scene. 
More recently, decoder-only Transformer models have become the mainstream in generative models. 
Represented by Codex~\cite{chen2021evaluating}, CodeGen~\cite{Nijkamp2022CodeGenAO}, and others like~\cite{Fried2022InCoderAG,Lu2021CodeXGLUEAM,Takerngsaksiri2022SyntaxAwareOC,Xu2022ASE,Zheng2023CodeGeeXAP}.

Nowadays, to Software Engineering developers, LLM-empowered code generation has become an irreplaceable assistant tool in their development cycle~\cite{vaithilingam2022expectation, JetBrains}. 
Although the code-LLM has been developed for years, those code generation tools still play a supporting role in the actual SE development.
Essentially, there are significant differences between the code generated by LLMs and that crafted by human developers.
Specifically, we conclude the phenomenon in the following threefold:
\begin{itemize} [leftmargin=*]
    \item \textbf{Observation 1:} Followed by the innate features of LLMs, the content and quality of generated codes heavily depend on the human-crafted prompts.

    \textbf{Discussion 1:} Providing longer prompts with richer content seems a plausible solution. With the growth of LLM size, the maximum input length increases from 512 tokens for the early models~\cite{Wang2021CodeT5IU} to 32,000 tokens for the extended GPT-4 models~\cite{GPT4}. However, it is still impossible to feed the entire code repository and all other relevant information into the LLMs. Furthermore, studies have demonstrated that irrelevant context in prompts not only affects model efficiency and increases expenses, but also potentially compromises the model performance~\cite{Shi2023LargeLM}.
    Therefore, to maintain the practicability and effectiveness simultaneously for code generation tasks, external context information should be extracted and selectively fed into LLMs.
    
    \item \textbf{Observation 2:} Existing methods focus on generating independent and often small-scale code units~\cite{du2023classeval, yu2023codereval}.
    According to a previous study~\cite{du2023classeval}, the code generated by LLMs shows lower method-level correctness. Although various benchmarks have been proposed (Details are discussed in the Related Work (Section~\ref{sec_related})), newly generated code remains questionable. 
    
    \textbf{Discussion 2:} The correctness relates to many factors, and evaluation of correctness is already challenging enough. Studies have shown that larger and more powerful foundation LLMs lead to better code generation correctness. 
    Additionally, retrieving and reusing existing functions or APIs (which are usually verified by other human developers), as the skeleton for code generation, could also fundamentally improve the correctness. This is also the strategy that human developers commonly apply in practice. 

    \item \textbf{Observation 3:} Some code-LLMs (e.g., GPT-4) have developed superior skills in simple algorithms such as greedy algorithm and searching strategies, but they perform poorly in complicated data structures such as tree structure and graph theory~\cite{chang2023survey}.

     \textbf{Discussion 3:} Similarly, instead of generating complex code snippets from scratch, importing functions that are already implemented can mitigate the shortcomings of advanced-level programming challenges. For example, Python developers are likely to import functions from \href{https://networkx.org/documentation/stable/index.html}{\textcolor{blue}{networkx}} package to build a graph, not building graphs by the dictionary-like data structure.
\end{itemize}

\subsection{Code Generation in Code Repository}
During software development, programmers typically work within a specific code environment, essentially a designated code repository. 
This repository encompasses all project modules, functions in these modules, and third-party libraries, serving as the foundational framework for building and expanding software applications. 
In this context, developers, especially experienced ones, must thoroughly navigate and understand this code repository environment to ensure seamless integration of newly added code with the existing one.
When it comes to LLM-based code generation techniques, solely relying on inputting developer requirements to these models fails to fully utilize the potential of LLMs in interacting with existing code repository environments.
Specifically, these techniques are blind to the repository’s unique naming conventions and programming styles, particularly when the completion target is relevant to cross-file context~\cite{Ding2022CoCoMICCC,Pei2023BetterCM, Shrivastava2023RepoFusionTC}.
Instead of using a single code file, well-designed programs emphasize separating complicated functionality into independent modules, each implementing only one aspect of the desired functionality~\cite{Quist2022ProceedingsOT,Turner1980TheSO}.

As illustrated in Fig.~\ref{fig:killing example}, in the \textit{Direct Generation} method, developers input their requirements directly into ChatGPT, expecting to receive code for a function that executes the task of \textit{``checking to see if a tag contains bulleted text''}.
However, the generated code, due to its lack of knowledge about local modules, fails to adequately consider a crucial factor: before determining whether a tag contains bulleted text, it's essential to first extract the text content from the provided \textit{tag\_elem}, which is a vital prerequisite.
As a result, the generated code (highlighted in the yellow box) simply searches for content within the \textit{$<$ul$>$} and \textit{$<$ol$>$} tags.
If bullet text exists outside these tags, for example, in the input \textit{``$<$div$>$\textbackslash u0095 This is a list item.$<$/div$>$''}, the generated code will be unable to extract text content, thus preventing accurate detection.

Furthermore, since the model is unaware of the existence of the ``\textit{check text for bullet characters}'' function in the global module, it defines a new function \textit{is\_bullet\_text} (highlighted in blue) in the generated code.
A large part of this code duplicates the function \textit{is\_bullet\_text} in the global module \textit{unstructured/partition/text\_type.py}, resulting in code redundancy. 
It also brings additional concerns about the correctness of the newly generated function.

Additionally, since the model is unaware of which third-party libraries are available, it arbitrarily imports and uses the bs4 (\href{https://pypi.org/project/beautifulsoup4/}{\textcolor{blue}{BeautifulSoup4}}) library. 
This can result in the generated code that refers to a library that is not available, leading to the bug ``\textit{ModuleNotFoundError: No module named ***}''.

These knowledge gaps highlight the limitations of existing code generation approaches, highlighting the need for more context-aware methods that can effectively interact with the code repository environment. 
In summary, a lack of knowledge about local and global modules can result in models generating code lacking functional constraints or containing redundancies.
A lack of knowledge about third-party libraries in the code repository can lead to models using arbitrary third-party libraries, causing problems with the generated code not working.
As a result, generating code in specific modules of the code repository is more than just meeting functional requirements. 
To generate code similar to human developers, the approach must fully understand and effectively utilize this diverse contextual knowledge to ensure that newly introduced code seamlessly coordinates with the existing code repository.

\subsection{Bridging the Gap: Making LLMs Aware of Three Types of Repository Knowledge}

Above all, in order to enable LLMs to fully utilize information spanning over the code repository environment for code generation, it is essential to provide the model with three types of knowledge: 
local modules, global modules, and third-party library information. 

As shown in Fig.~\ref{fig:killing example} \textit{Local Module Aware \& Global Module Aware \& Third-Party Library Aware Code Generation}, when the developer requirement is entered, since the model is aware of the knowledge of the local module, the generated code reuses the \textit{\_construct\_text} function (highlighted in yellow) in the local module to extract the text content in the text label element.
In addition, because the model is aware of knowledge of the global module, the generated code reuses the \textit{is\_bulleted\_text} function (highlighted in blue) from the global module \textit{unstructured/partition/text\_type.py}, thus reducing redundant code definitions. 
In addition, the model is aware of the existence of third-party libraries and therefore reuses the \textit{lxml} library to ensure its availability.

In the next section, we comprehensively describe how to extract, structure, and utilize local aware, global aware, and third-party library aware information from code repositories.

\begin{figure*}%[H]
    \centering
    \includegraphics[width=1\textwidth]{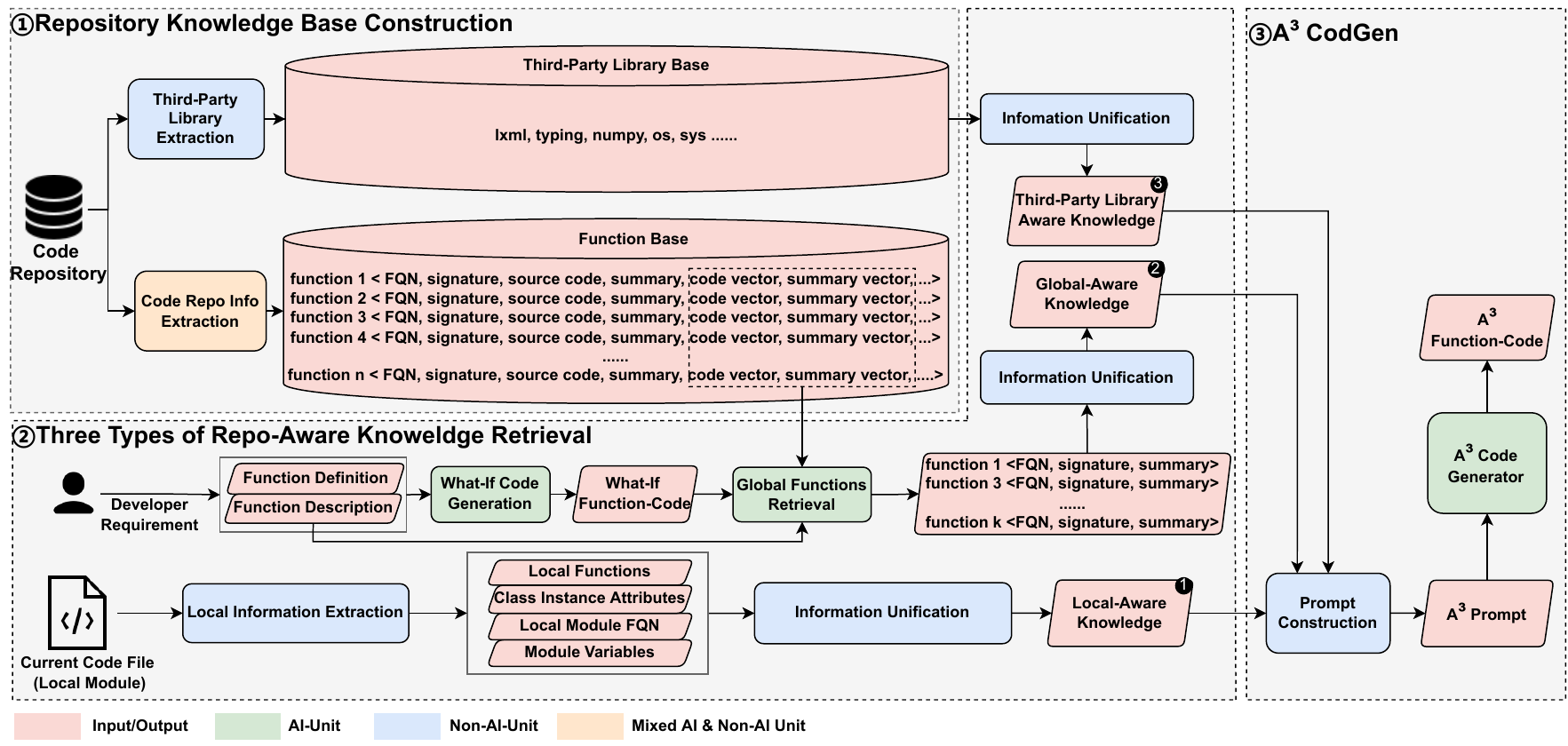}
    \caption{$A^3$-CodGen Framework.} 
    \label{fig:approach overview}

\end{figure*}

\section{\texorpdfstring{$A^3$}-CodGen Framework} \label{sec:AAA code generation framework}
In this section, we elaborate on our novel $A^3$-CodGen Framework to harness the full potential of code repository knowledge for assisting LLMs in code generation.
As depicted in Fig.~\ref{fig:approach overview}, the $A^3$-CodGen Framework comprises three essential phases.
First, we obtain the function base and third-party library base (Section~\ref{sec:code repo extract}),  providing sufficient knowledge to support subsequent retrieval. 
Then, we retrieve the code repository knowledge from other code files, the third-party library repository, and the current file. 
After that, we organize them into 3 types of code repository aware knowledge (Section~\ref{sec:three aware knowledge mining}).
Finally, the three code repository-aware knowledge are fused into a prompt (Section~\ref{sec:three aware knoweldge utilization}), serving as input to the foundation LLM, to ultimately generate codes with three essential awareness.

\subsection{Repository Knowledge Base Construction} \label{sec:code repo extract}
Code repositories contain a wealth of knowledge to exploit, but this knowledge usually exists in an unstructured form and cannot be directly utilized effectively, scattering in the code repository. 
Therefore, we need to perform systematic knowledge extraction to transform and compress this knowledge into a structured form in order to build a knowledge base to support subsequent knowledge utilization, such as knowledge retrieval.
Thus, we perform \textbf{code repository information extraction} (Section~\ref{sec:repo info extraction}) and \textbf{third-party library extraction} (Section~\ref{sec:third-party lib extraction}) to obtain the function base and third-party-library base, respectively.

\subsubsection{Code Repository Information Extraction} \label{sec:repo info extraction}
For a code repository, we can effortlessly obtain all code files in the entire repository by traversing all the directories within the project.
Then, for each file, we extract all included functions, gathering both basic code information and semantic code knowledge.
This process snowballs to extract all essential repository information.

\textbf{Basic Code Knowledge Extraction.}
Abstract Syntax Tree (AST) is a tree representation of the source code, and studies have demonstrated its usefulness on various SE tasks~\cite{zhang2019novel}.
In this study, we employ the Python library \href{https://docs.python.org/3/library/ast.html}{\textcolor{blue}{ast}} to analyze the AST for each source code file in the code repository.
From the AST, we extract six essential types of basic information:
1) Comment.  
2) Signature. 
This knowledge illustrates the parameters and returns of a function.
For example, the signature of the function \textit{\_construct\_text} is \textit{def \_construct\_text(tag\_elem, include\_tail\_text)-$>$str}.
3) Function Source code.
4) Class associated with the function (if a function is a member of a class). 
5) Is\_Empty tag. 
\textit{True} means empty function, \textit{False} means empty label.
For example, function \textit{\_construct\_text} has \textit{is\_Empty} value \textit{False}.
6) The file path of the function. 
This reflects the relative location of the current function in the entire code repository.

\textbf{Semantic Code Knowledge Extraction.}
AST is criticized for its prone to long-term dependency problems. 
Also, simply employing AST potential weakens the capability to capture real and complex semantics~\cite {zhang2019novel}.
Thus, we extract three types of semantic code knowledge to enhance the basic code knowledge, as follows:

1) Fully Qualified Name (FQN) of Functions.
FQN indicates the function's origin in the code repository and, if applicable, the specific class within the module.
For example, in the local module of the Fig.~\ref{fig:killing example}, the function \textit{\_construct\_text} (highlighted in yellow) has the FQN \textit{unstructured.documents.html.\_construct\_text}.

2) Function Summary.
Not every function is elaborated by comments.
Even comments exist, their quality varies;
some lack clear descriptions of the function's basic functionality, while others might be over-detailed. 
Overly lengthy descriptions can distract from understanding the function's actual usage and behavior.
Therefore, we generate a function summary to precisely describe the purpose and functionality of the function.
Previous work has demonstrated that large language models (LLMs) are capable of in-context learning~\cite{Huang2022SEFK,Brown2020LanguageMA,ye2023context}, where a task description and examples can enable LLMs to perform well. 
Thus, we use GPT-3.5-Turbo~\cite{GPT-3.5-Turbo-16k} as the foundation model due to its demonstrated code understanding and summarization capabilities, as well as its cost-effectiveness. The selection of the foundation model is flexible, and our method is also applicable to advanced models such as GPT-4~\cite{GPT4}, although they come at a higher cost.

\begin{figure}%[H]
    \centering
    \includegraphics[width=0.48\textwidth]{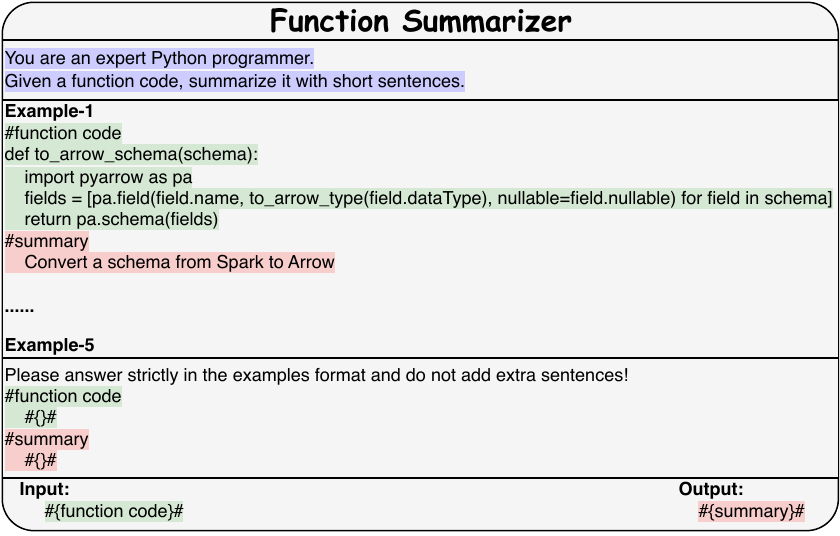}

    \caption{Function Summarizer Unit} 
    \label{fig_function_summarizer}
    \vspace{-3mm}
\end{figure}

\textbf{Function Summarizer Unit.}
To generate natural language semantic information for functions, we design a prompt as the core of the \textit{Function Summarizer Unit}. 
This prompt creates summaries for each function. 
As shown in Fig.~\ref{fig_function_summarizer}, the prompt includes a task description (highlighted in purple), five demonstration examples, and an input template.
The input to this unit is the function's code (highlighted in green), and the output is the model-generated summary (highlighted in red).

3) Vector Representations.
In the previous two steps, we underlined 
explictly select information.
To ensure efficient retrieval of global functions later on, having an effective code representation that faithfully captures function semantics is essential.
Hence, we employ a dual-pronged approach, conducting separate embeddings for both the source code and the function summaries.
This results in the creation of \textit{code vector} and \textit{summary vector}, respectively.

\subsubsection{Third-Party Library Extraction} \label{sec:third-party lib extraction}
Third-party libraries often appear in code files, and when specific libraries are provided to a model, it likely has learned how to use them during training~\cite{Huang2023APIEA, Huang2022PrompttunedCL}. 
Therefore, supplying just the library name allows the model to automatically understand how to use it for a given task. 
For example, we can ask the model to use the NumPy library to compute the cosine similarity, given the input \textit{vector\_a} and \textit{vector\_b}.
In this case, the model can directly call functions in the NumPy library, such as \textit{numpy.dot} and \textit{numpy.linalg.norm}, to perform the required computations.
The pre-trained model's prior knowledge and extensive large-scale training enable them to effectively utilize third-party libraries, improving the efficiency and accuracy of code generation.
Therefore, we extract the import statements of third-party libraries directly from all code files via AST and saved their library names to construct the third-party library base.
It is important to note that the same library can have different versions, and API call methods may vary between versions.
To address this, the current repository's Python interpreter or the ``pip list'' command can be used to determine the versions of pre-installed libraries, guiding the model in generating accurate code.

\subsection{Three Types of Repo-Aware Knowledge Retrieval} \label{sec:three aware knowledge mining}
As we discussed in Section~\ref{sec_motivation_limitation}, although the maximum input length grows greatly, it is still impossible to feed the entire code repo and third-party library information to LLMs.
Furthermore, irrelevant context not only affects model efficiency and increases expenses, but also potentially compromises the model performance~\cite{Shi2023LargeLM}.

To enable the model to fully utilize the three types of information extracted from the code repository, we conduct comprehensive knowledge mining: 
extracting four types of \textbf{local information} (Section~\ref{sec:local-module knowledge}), retrieving global functions from the function base using developer requirements to obtain \textbf{global-aware information} (Section~\ref{sec:global-module knowledge}), and structuring third-party library information from the \textbf{third-party library base} (Section~\ref{sec:third-party lib knowledge}).

\subsubsection{Local-Module Knowledge Mining} \label{sec:local-module knowledge}
Within the user's working context, referred to as the local module, four types of local-module information can be encountered:

\begin{itemize} [leftmargin=*]
    \item \textbf{Local functions.}
    Local functions encompass all functions that are defined and implemented within the local module. 
    We extract these local functions and represent them in the form of $<$Fully Qualified Name (FQN), Function Summary, Function Signature$>$. 
    The FQN provides details about the function's location within the module, the Function Summary reflects the natural language description of the function's purpose, and the Function Signature reveals its input and output specifications.
    
    \item \textbf{Class Instance Attributes.}
    Class instance attributes (or instance variables) are variables defined within a class's constructor function (\textit{\_\_init\_\_} in Python).
    These variables are specific to instances of the class and store instance-specific attributes or state information.
    Some local functions may be defined within a class might involve interactions with class instance attributes, accessing or modifying them to achieve specific functionality.    
    \item \textbf{Local Module FQN.}
    Local Module FQN partially reflects the structural information of the working context (local module) within the entire code repository.
    It helps the model understand the context of how local functions, class instance attributes, and variables within this module are used. 
    
    \item \textbf{Module variables.}
    Local modules may define variables, some of which can directly interact with local functions, potentially supporting the functionality of newly added functions.
    We unify the declarations of these variables. However, if there are no variables defined within the module, we indicate this with \textit{None}.

\end{itemize}

\subsubsection{Global-Module Knowledge Mining} \label{sec:global-module knowledge}
In addition to using local module knowledge, global module knowledge also plays a significant role in code generation.
Instead of letting the model implement complicated logic steps from scratch, we expect the model can directly call functions if they are already present in other code files.
To help models realize that global functions can be reused, we retrieve global functions in the code repository and organize them in a structured way.
The global module knowledge mining process is divided into three phases: 
what-if code generation, global function retrieval, and information unification.

% \begin{figure}[t]
%     \centering
%     \includegraphics[width=0.47\textwidth]{A3_CodeGen_AIUnit.pdf}

%     \caption{$A^3$ Code Generator Unit} 
%     \label{fig:Code Gen Units}

% \end{figure}

\begin{itemize} [leftmargin=*]

\item \textbf{What-If Code Generator Unit:}
In Section~\ref{sec:repo info extraction}, we construct a function base derived from the code repository. 
To retrieve global functions from this base, we consider both textual and code semantic similarity.
For code semantic similarity matching, we utilize the code generation capability of the LLM, inputting user requirements to generate a \textit{what-if code}, which encapsulates the logic to fulfill those requirements and serves as input for code similarity matching in the Global Function Retrieval Unit.
For in-context learning, we provide three representative examples. 
Notably, as the generated code is not the final code and only used to retrieve the global function, so completeness is not required.
As shown in the experiment results in Section~\ref{sec:RQ2 topk}, using function descriptions for text similarity and what-if code for code similarity retrieval can retrieve more useful global functions compared to using only function descriptions.

\item \textbf{Global Function Retrieval Unit}: This unit uses function descriptions and the what-if function-code as input to retrieve reusable global functions from the function base.
Specifically, we first embed \textit{function\_description} into $vector_{des}$ and compute the cosine similarity between this vector and the summary vectors of all functions in the function base. 
Thus, we can rank and obtain the top k $RetrievedFun_{des}$.
Second, we embed the what-if function-code as a $vector_f$ and compute the cosine similarity between this vector and the code vectors of all the functions in the function base, so we can rank and obtain the top k $RetrievedFun_{code}$. 
Finally, we merge the sets of $RetrievedFun_{des}$ and $RetrievedFun_{des}$ to obtain the global function that is most relevant to the provided input, which is denoted as $RetrievedFun$.

\item \textbf{Information Unification Unit}: For each $RetrievedFun$, we organize and display the retrieved information as $<$FQN, function signature, summary$>$ in a structured format.
\end{itemize}

\subsubsection{Third-Party Library Knowledge Mining} \label{sec:third-party lib knowledge}
In the process of code generation, to behavior similar to human developers, it's crucial for the LLM to be aware of the third-party libraries pre-installed in the current code repository environment.
Without this awareness, there's a risk of generating code that relies on third-party libraries unsupported by the current environment, rendering the code non-functional.
To ensure that, we create a comprehensive list of all the third-party libraries found in the third-party library base, as extracted in section~\ref{sec:third-party lib extraction}. 
It's important to note that we extract only the names of these libraries and do not include any of their code. 
This approach is supported by recent studies showing that LLMs can serve as neural knowledge bases for SE knowledge due to their pre-training on extensive source code corpora, which includes third-party library knowledge~\cite{Zan2022CERTCP, Liu2023CodeGen4LibsAT, Ma2024CompositionalAR, Huang2022SEFK}.
Several works demonstrate library-oriented code generation, where LLMs generate code based on specific library APIs~\cite{Zan2022CERTCP, Liu2023CodeGen4LibsAT, Ma2024CompositionalAR, Ren2023FromMT}.
By constructing this list of third-party libraries, we empower the LLM to generate code that aligns with the available environment and utilizes the appropriate libraries effectively, reducing the risk of non-executable code.

\subsection{\texorpdfstring{$A^3$} \ Code Generation} \label{sec:three aware knoweldge utilization}

After section~\ref{sec:three aware knowledge mining}, we obtain structured local-aware knowledge, global-aware knowledge, and third-party-library-aware knowledge. 
To ensure the model's full awareness of these three types of knowledge, we design an \textbf{$\mathbf{A^3}$ Code Generation Unit} that seamlessly integrates user requirements with the three knowledge sources to generate $A^3$ functional code.

\begin{figure}[t]
    \centering
    \includegraphics[width=0.47\textwidth]{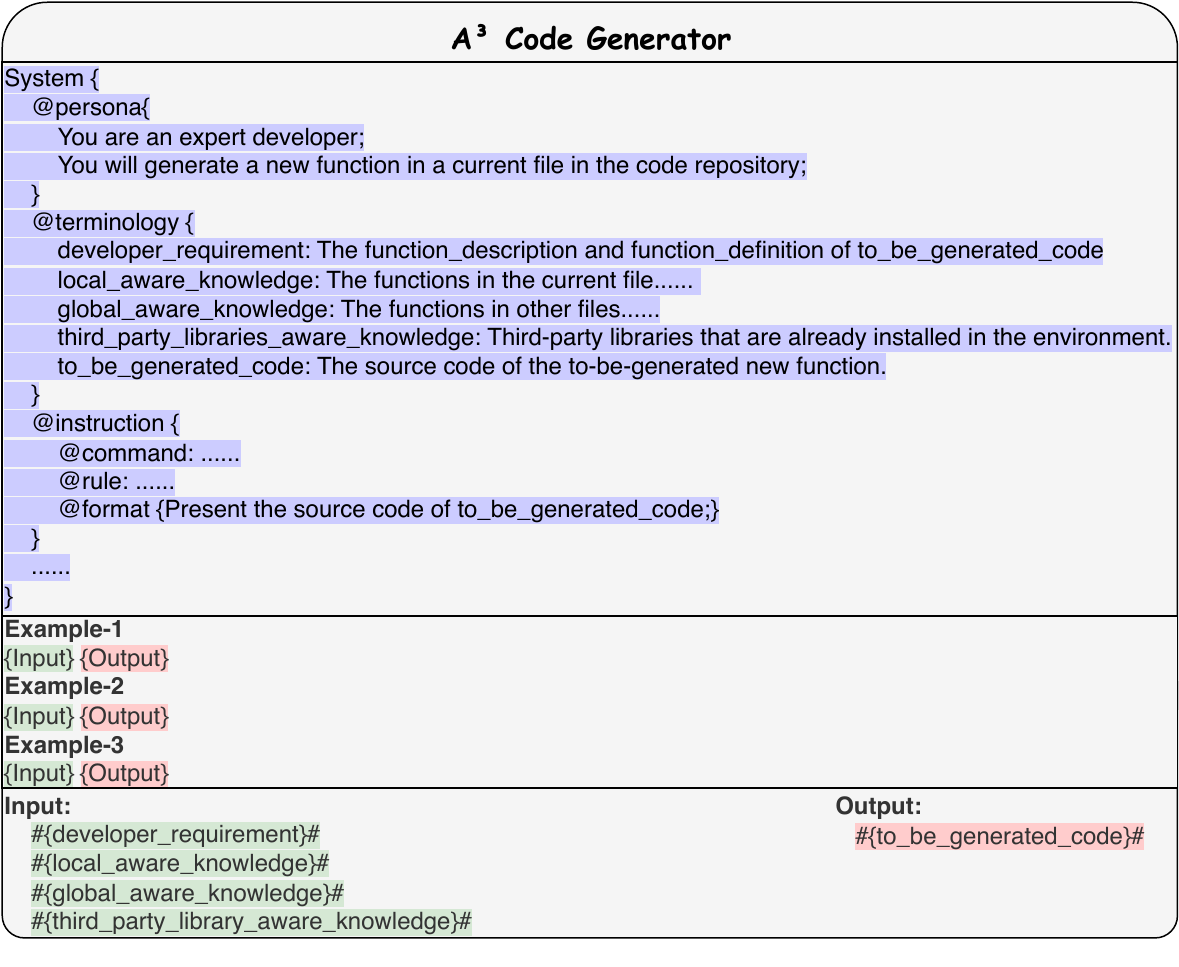}

    \caption{$A^3$ Code Generator Unit} 
    \label{fig:Code Gen Units}
    \vspace{-3mm}

\end{figure}

The $A^3$ Code Generation Unit, as depicted in Fig.~\ref{fig:Code Gen Units}, consists of four key components: 
task description, examples, input, and output.
In the task description (purple part), we employ a structured prompt to delineate the repository-aware code generation task. 
Key components of the prompt include:
1) @persona: Guiding the model to assume the role of an expert developer, providing clarity on the task.
2) @terminology: Defining crucial information in a keyword format.
3) @instruction: This section furnishes explicit instructions for executing the task:

\begin{itemize} [leftmargin=*]
    \item @command: Offering an overarching strategy for the model to follow during the task.
    \item @rule: Outlining the basic limitations and considerations to be observed during code generation.
    \item @format: Detailing the expected presentation format for the source code generated for the new function.
\end{itemize}

In the @command, we guide the model through the code generation process step by step by employing chain-of-thought~\cite{wei2022chain,wang2022self,lyu2023faithful}.
\begin{itemize} [leftmargin=*]

    \item Firstly, in @command1, we prompt the model to contemplate the logic and steps required to fulfill the developer's request.
    \item Next, in @command2, we encourage the model to consider whether there is any reusable information within the local-aware knowledge and utilize it if available.
    \item Following that, in @command3, we guide the model to think about whether there are any reusable global functions within the global-aware knowledge and utilize them if present.
    \item Subsequently, in @command4, we prompt the model to explore how to leverage the third-party library-aware knowledge to interact with libraries effectively.
    \item Finally, in @command5, if the model cannot leverage any information, it is instructed to generate code directly based on the logic and steps it has formulated.
    \item Additionally, we include an @command that reiterates the importance of strictly following the steps from 1 to 5.
\end{itemize}

In the @rule, we establish three rules to govern the code generated by the model, including calling import FQN, restricting the use of specified third-party libraries, and ensuring that the generated code is not empty (as the model may occasionally generate abstract functions).
We then provide three representative examples.
Finally, we input the developer requirement and our 3 knowledge into Unit to get the $A^3$ function code (i.e., \_to\_be\_generated\_code in Fig.~\ref{fig:Code Gen Units} (b)).
Notably, developers only needs to input requirements, and all 3 types of repo-aware knowledge are obtained automatically.

Due to the page limitation, we have organized all prompts for our AI units (e.g., Code Summarizer, and $A^3$ Code Generator) in the ``prompts/'' folder of our artifact~\cite{our_repo}.  
\section{Experiment Setting}

In this section, we first outline the selection of foundation models, the dataset, and evaluation metrics for assessing our framework.
We then introduce five Research Questions (RQs):
\begin{itemize} [leftmargin=*]
    \item RQ1: Does Local Module Aware Knowledge indeed help LLMs fully utilize local module knowledge during code generation, and which types are most effective in improving local awareness?

    \item RQ2: Is Global Module Aware Knowledge effective in assisting LLMs to utilize global module knowledge during code generation, and what is the optimal ``k'' value for retrieving global functions?
    
    \item RQ3: Does Third-Party Library Aware Knowledge significantly impact LLMs' awareness of available third-party libraries during code generation?
    \item RQ4: What is the overall performance of our $A^3$-CodGen framework?
    \item RQ5: How does $A^3$-CodGen framework compare to existing code generation tools like GitHub Copilot?
\end{itemize}

\subsection{Foundation Models}
We select GPT-3.5-Turbo-16k~\cite{GPT-3.5-Turbo-16k} as the default foundation LLM for this study due to its excellent performance in various SE tasks, along with its popularity and cost-efficiency.
As for the embedding model, we use the \textit{ada-embedding-002} model~\cite{text-embedding-ada-002}, which generates high quality sentence embeddings for effective similarity computation and retrieval operations.~\cite{Gao2023Vec2VecAC,Gnther2023JinaE2}.
Similar to other LLM-empowered frameworks in SE domain, the more powerful model is selected, the better performance can be achieved, generally.

\subsection{Dataset Collection}
While there exist some benchmark datasets designed for code generation, such as~\cite{yuan2023evaluating, liu2023your, du2023classeval, ouyang2023llm, yu2023codereval}, 
they suffer from the following issues: 
1) some of the existing datasets contain code repositories that are already present in the training corpora of popular LLMs, leading to an overestimation of the code repository-aware code generation capabilities~\cite{Matton2024OnLO, Yang2023UnveilingMI}. 
2) existing benchmarks only evaluate class-level or small unit test cases, which is not suitable for our research goals.
3) Based on our literature review, none of existing code generation benchmark datasets can evaluate whether $A^3$-CodGen can effectively obtain, understand, and correctly reuse three distinct types of code repository information. 
Therefore, to address the potential risk of data leakage and accurately reflect the capability of $A^3$-CodGen, we created a new benchmark dataset, dubbed \textbf{RepoEval}. 
This decision stems from our intention to verify that the model truly grasp $A^3$ code generation after receiving information we provide.

%, rather than relying solely on memorized code patterns seen during their previous training.

\subsubsection{Code Repository Collection} \label{sec:code repo collection}

To construct a benchmark dataset, we follow methodologies similar to those in previous works~\cite{Tufano2022UsingPM, Mastropaolo2022UsingDL, Alfadel2021EmpiricalAO}. 
As we focus on Python code repositories, it is common to collect packages from the \href{https://pypi.org/search/?c=Programming+Language+%3A%3A+Python+%3A%3A+3&o=-created&q=&page=1}{\textcolor{blue}{PyPI}} ecosystem, as mentioned in Alfadel et al.~\cite{Alfadel2021EmpiricalAO}.

To ensure that the chosen repositories are not part of the training data for the models used, we verify the release dates of each repository to identify their first release date. 
If the first release date is after the training period of the model, it means that the repository was not included in the training data.
According to \href{https://platform.openai.com/docs/models/gpt-3-5}{\textcolor{blue}{OpenAI's official website}}, the training data for GPT-3.5-Turbo-16k is cut off as of September 2021. 
Therefore, we select projects with the earliest release date after October 1, 2021, from PyPI.
We further filter the GitHub repositories by choosing those with more than 100 stars and 100 branches, as these metrics indicate repository quality~\cite{Mastropaolo2022UsingDL}.
Ultimately, we identify 29 code repositories to construct our benchmark, each averaging 82.517 code files.

Then, we create 29 function bases and 29 third-party library bases for the corresponding code repositories, following the method outlined in Section~\ref{sec:code repo extract}.
The function bases from these 29 code repositories encompass a total of 13,784 functions.
Out of these, 10,077 are class member functions, leaving 3,707 as module functions.
Since class member functions often involve complex relationships related to inheritance and other classes, it becomes highly labor-intensive to analyze the code invocation relationships to construct the benchmark.
Therefore, our benchmark dataset focuses solely on module functions.
From these 3,707 module functions, we filter out functions that had comments and were non-empty, and finally, we obtain 1,533 functions.

\subsubsection{Benchmark Construction} \label{sec_benchmark_construction}
Given the comments and the definition of a function, the corresponding source code for that function serves as the standard answer to be generated. Simultaneously, we extract call information for the function to determine the extent of code reuse within it, because a higher extent of function reuse reflects the framework learns better from three awareness.

Given a function, code reuse falls into one of the three categories:
1) It could be part of locally defined functions within the local module;
2) belong to a global module; or
3) belong to a third-party library.
To obtain this reuse information, we need to analyze the call relationships between functions among codes. 
However, existing code analysis tools like pycallgraph~\cite{pycallgraph} require code compilability to accurately generate call graphs for functions.
Considering that it is impractical to set up running environments for all these code repositories due to diverse requirements and potential conflicts, and given that static analysis tools like pyan~\cite{pyan} have relatively lower accuracy and cannot meet our needs, we resort to manual annotation to extract the call information between functions.

Manual annotation for each function is resource-intensive and time-consuming. 
To ensure experiment efficiency, we employ a statistical sampling method~\cite{Singh1996ElementsOS}. 
Specifically, we perform sampling on the 1,533 module functions obtained in Section~\ref{sec:code repo collection}. 
We ensure that the estimated accuracy within the sample remains within a 0.05 error range at a 95\% confidence level while ensuring functions from each code repository are included in the sample.
Finally, we acquire 383 functions for subsequent experiments.

\subsection{Metrics} \label{sec_metrics}
To comprehensively evaluate whether the three awareness are effective, we employ the following metrics:

\begin{itemize} [leftmargin=*]

    \item  \textbf{Reuse Awareness}: To assess the model's ability to correctly reuse local, global, and third-party library information when necessary and avoid reuse when unnecessary, we computed the following awareness metrics:
    \begin{itemize}
        \item \textbf{True Positive (TP)}: The model-generated code correctly reuses local, global, or third-party library information, consistent with the benchmark code. 
        For instance, if the model-generated code reuses local functions and the benchmark code also reuses local functions, it is considered TP in the case of local reuse, as well as in the case of global reuse and third-party library reuse.
        \item \textbf{False Positive (FP)}: The model-generated code incorrectly reuses local, global, or third-party library information, where no reuse occurs in the benchmark code.
        \item \textbf{False Negative (FN)}: The model-generated code fails to reuse local, global, or third-party library information, but reuse occurs in the benchmark code.
        \item \textbf{True Negative (TN)}: The model-generated code correctly does not reuse local, global, or third-party library information, consistent with the benchmark code.
        \item \textbf{Precision}: Calculated as $TP/(TP+FP)$, indicating the proportion of the model-generated code that is correctly reused.
        \item \textbf{Recall}: Calculated as $TP/(TP+FN)$, indicating the proportion of the code that actually needed to be reused and is correctly reused by the model.
        \item \textbf{F1}: The harmonic mean of Precision and Recall, providing an overall measure of the model's effectiveness in reusing code, calculated as $2*(Precision*Recall)/(Precision+Recall)$.
        \item \textbf{Accuracy}: Calculated as $(TP+TN)/(TP+FP+TN+FN)$, representing the percentage of cases where the model is correctly aware or unaware.
    \end{itemize}
    \item \textbf{Reuse Correctness}: 
    After identifying local/global/third-party library information, we assess the correctness of reuse.
    The metric is determined by the number of exact matches between the functions/libraries reused by the model and those in the benchmark, divided by the total number of functions/libraries reused by the model. 
    For instance, in local reuse, if the model reuses functions $local_a$, $local_b$, and $local_c$, and the benchmark reuses functions $local_b$, $local_c$, and $local_d$, there are 2 exact matches, resulting in a correctness score of 2/3 = 0.667.
    It's important to note that we do not use conventional metrics like BLEU~\cite{Papineni2002BleuAM}, ROUGE~\cite{Lin2004ROUGEAP}, or METEOR~\cite{Banerjee2005METEORAA}, as they only evaluate token-level similarity between the model-generated code and the benchmark code. These metrics do not effectively measure the model's ability to correctly reuse code components during generation.

    \item \textbf{Library Coverage:}
    This metric represents the proportion of libraries used in the generated code that are already pre-installed in the code repository's environment.

    \item \textbf{Lines of Code (LOC):} This metric reflects code conciseness, determined by counting lines after removing comments. Lower LOC indicates more concise code.

\end{itemize}

To validate the call relationships between the extracted functions, we recruit two developers with over five years of Python programming experience.
Each function in the benchmark dataset is independently labeled by both developers.
They evaluate whether the function utilizes local functions within the current code file, assigning a score of 1 if it does and 0 if it doesn't. 
Similarly, they assess whether the function utilizes global functions and third-party libraries, assigning scores of 1 or 0 accordingly. 
These scores contribute to calculating reuse awareness.
For functions involving reuse, they identify all reused functions/libraries to calculate reuse correctness.

After the independent labeling, discrepancies are discussed to reach a consensus.
If any disagreements remain, the first author resolves them and makes the final decision.
In subsequent experiments, we use this benchmark for evaluation and annotate the model-generated code using the same method.
To label the model-generated results, we recruit 4 master's students and 8 undergraduate students, each with at least 2 years of Python programming experience. 
After a 1-hour training session on our standardized method, tasks are evenly distributed among them.
We utilize Cohen's kappa~\cite{Landis1977AnAO} to assess inter-rater agreement, with coefficients ranging from 86\% to 95\% for all annotated data (including the benchmark dataset and model-generated results), indicating high agreement among assessments.

\section{Evaluation}
%\sd{In addition to explanation of tables, it might provide more insightful knowledge if we can connect observations to some conclusions from previous empirical/user studies (such as Tingting's works). Just a suggestion.}
%\sd{Add a comparison for only NL input and NL+function input.}

% \subsection{RQ1: The Effectiveness of Local-Aware Knowledge}
% \label{sec_evalutaion_RQ1}

\subsection{RQ1: Does Local Module Aware Knowledge indeed help LLMs fully utilize local module knowledge during code generation, and which types are most effective in improving local awareness?}
\label{sec_evalutaion_RQ1}

\subsubsection{Motivation}
Local information aids LLMs in code generation by offering awareness of local contextual details. 
We extract and organize four types of Local-Aware Knowledge from the local module:
local functions, class instance attributes, local module FQNs, and module variables.
These knowledge types offer diverse perspectives on the context of the local module.
Our goal is to assess whether local-aware knowledge genuinely enhances LLMs' ability to reuse local functions in code generation.
Additionally, we aim to identify the optimal configuration, determining whether all four knowledge categories should be available to the model's awareness or only specific ones.

\subsubsection{Experiment Methodology} \label{sec:RQ1 methodology}
We use vanilla GPT-3.5-Turbo-16k as a baseline, which does not have any local information in its input. 
We define four variants of local-aware methods, as shown in Table~\ref{table:RQ1 local info}.

\begin{itemize} [leftmargin=*]
    \item \textit{+Local Functions} signifies the use of Local Functions knowledge from the current file context.
    \item \textit{+Class Instance Attributes} indicates the use of Class Instance Attributes Knoweldge
    \item \textit{+Module Variables} involves using the module Variables knowledge in the current file context.
    \item \textit{+Local Module FQN} employs the fully qualified name corresponding to the current file.
\end{itemize}

We run these five different configurations on the benchmark to obtain code generation results.

\begin{table}[t]
    \centering
    \caption{Comparison of the Local Reuse Awareness results obtained with different Local Information.
    % And determine the best configuration for Local Information Extraction.
    }

    \label{table:RQ1 local info}
    \resizebox{0.46\textwidth}{!}{
    \begin{tabular}{|l|c|c|c|c|}
        \hline \makecell[c]{Variant} &Precision & Recall & F1 & Accuracy\\ \hline
        Vallina GPT-3.5-Turbo-16k & N/A & N/A & N/A & 0.721 \\ \hline        
        \makecell[l]{GPT-3.5-Turbo-16k\\ +Local Functions} & 0.667 & 0.528 & 0.589 & 0.796 \\ \hline
        \makecell[l]{GPT-3.5-Turbo-16k\\ +Local Functions\\ +Class Instance Attributes} & 0.720 & \textbf{0.632} & \textbf{0.673} & \textbf{0.830} \\ \hline

        % \makecell[l]{ChatGPT\\+Local Functions\\+Class Instance Attributes\\+Local Module FQN} & 40 & 55 & 263 & 26 & 0.4211 & 0.6061 & 0.4969 & 0.7891\\ \hline

        \makecell[l]{GPT-3.5-Turbo-16k\\ +Local Functions\\ +Class Instance Attributes\\ +Module Variables} & 0.728 & 0.557  &  0.631 & 0.820 \\ \hline
        
        \makecell[l]{GPT-3.5-Turbo-16k\\ +Local Functions\\ +Class Instance Attributes\\ +Module Variables\\ +Local Module FQN} & \textbf{0.759} & 0.566 & 0.649 & 0.830 \\ \hline

    \end{tabular}
    }

\end{table}

%取平均值，提升的数值加百分比%，加粗

%这个RQ看完成任务的基础上，还能够精良复用code repo代码

\subsubsection{Result Analysis} \label{sec:RQ1 result}
Table~\ref{table:RQ1 local info} shows the reuse awareness results for code generation using vanilla GPT-3.5-Turbo-16k and four types of Local-Aware Knowledge Configurations.
Vanilla GPT-3.5-Turbo-16k exhibits a precision and recall of 0 as it does not recognize any local functions.
Consequently, in its generated code, new functions are often defined to complete certain logic, resulting in code redundancy.

When given knowledge of \textit{Local Functions}, the model achieves a precision of 0.667, recall of 0.528, F1 score of 0.589, and accuracy of 0.796.
This is because \textit{Local Function} knowledge allows the model to directly invoke these functions.
Upon additionally providing knowledge of \textit{Class Instance Attributes}, the model's precision, recall, and F1 score improve by 7.94\%, 19.70\%, and 14.26\%, respectively, compared to providing only knowledge of Local Functions. 
This improvement is attributed to the provision of \textit{Class Instance Attributes}, which enables the model to better understand how local functions interact with class variables in the project, leading to more accurate application of local functions.

However, when we introduce the \textit{Module Variables}, the model's performance drastically declines, particularly with a 11.87\% drop in precision.
This may be due to the increased information load resulting from knowledge of \textit{Module Variables}, which impacts performance. 
Nevertheless, the model's performance is restored and approaches the capabilities of providing knowledge of \textit{Local Functions} and \textit{Class Instance Attributes} when we subsequently add knowledge of Local Module FQN. 
This suggests the existence of subtle interactions within the knowledge configuration.
When the model receives only one or two key knowledge types, its performance may be superior to when it receives more information.
However, as the volume of knowledge increases to a certain extent, the model's capabilities strengthen, resulting in improved performance.

We ultimately choose the Local-Aware Knowledge setup, including knowledge of \textit{Local Functions} and \textit{Class Instance Attributes}.
This configuration serves as default for the upcoming experiments.

\begin{tcolorbox}

\textbf{Finding 1:} \emph{The Local-Aware Knowledge significantly enhances the model's ability to reuse local functions, thereby reducing code redundancy. 
The best configuration is the combination of \textit{Local Functions} and\textit{ Class Instance Attributes}.}

\end{tcolorbox}

\subsection{RQ2: Is Global Module Aware Knowledge effective in assisting LLMs to utilize global module knowledge during code generation, and what is the optimal ``k'' value for retrieving global functions?}
\label{sec:RQ2 topk}

\subsubsection{Motivation}
In this RQ, our goal is to explore the effectiveness of global-aware information in aiding LLMs to reuse global functions in code generation.  
When using global-aware knowledge, a critical issue arises:
if too few candidate functions are retrieved, the model may lack the necessary functions to generate code that correctly reuses global functions. 
Conversely, retrieving too many candidate functions can lead to the presence of irrelevant functions, making it difficult for the model to identify the required ones. Research shows that irrelevant context can decrease model efficiency, increase costs, and potentially compromise performance~\cite{Shi2023LargeLM}.
Therefore, we aim to identify the optimal top-k value for retrieving candidate functions to achieve the best global-aware configuration.
Additionally, we conduct an ablation study to evaluate the effectiveness of the What-if Code Generator Unit based on the optimal k value.

\subsubsection{Experiment Methodology}
We employ \textit{No Global-Aware} as the baseline and this involves providing only Local-Aware Knowledge without Global-Aware Knowledge input. 
To determine the optimal top-k value, we pre-define four variants: k=1, 5, 10, and 15.
These values are chosen because larger projects with more complex call relationships require at least one global function. 
A single retrieved global function may not be sufficient, and some functions may need to call multiple global functions, so we increase the k value to 5, 10, and 15.
This selection ensures we cover a range of potential requirements, balancing between too few and too many candidate functions to maintain model performance and relevance.
We run the total five different configurations on the benchmark to obtain code generation results. 
Subsequently, we assess the generated code using the same method as in section~\ref{sec_evalutaion_RQ1}, where each result is evaluated for whether it uses global functions.

To further evaluate the effectiveness of our \textit{What-if Code Generator Unit}, we conduct an experiment using the benchmark, focusing on test cases that call global functions (47 cases in total). 
We compare two retrieval methods:
\begin{itemize} [leftmargin=*]
    \item FunDes: Uses only function descriptions for text similarity retrieval.
    \item FunDes+FunCode: Uses function descriptions for text similarity retrieval and the What-If Code Generator to generate intermediate code snippets for code similarity retrieval.
\end{itemize}

%%----------------------------
%
%
\begin{table}[t]
    \centering
    \caption{Comparison of the Global Reuse Awareness results obtained for different Top k numbers of retrieved functions to find the best k value.
    % Where Precision, Recall, and F1 are for global functions.
    }

    \label{table:RQ2 repo info}
    \resizebox{0.47\textwidth}{!}{
    \begin{tabular}{|c|c|c|c|c|c|}
        \hline \makecell[c]{Global\\Functions of\\Different\\Top k} & \makecell[c]{Avg.\\Retrieved\\Functions} & Precision & Recall & F1 & Accuracy\\ \hline
        \makecell[c]{No Global-Aware\\(k = 0)} & N/A & N/A & N/A & N/A & 0.843 \\ \hline       
        k=1  & 1.757 & 0.372 & 0.533 & 0.438 & 0.786 \\ \hline        
        k=5 & 8.154 & \textbf{0.518} & \textbf{0.717} & \textbf{0.601} & \textbf{0.851} \\ \hline    
        k=10 & 16.031 & 0.479 & 0.583 & 0.526 & 0.836\\ \hline    
        k=15 & 23.592 & 0.407 & 0.583 & 0.479 & 0.802 \\ \hline    
    \end{tabular}
    }%

\end{table}

%
%
%%----------------------------

\subsubsection{Result Analysis}

Table~\ref{table:RQ2 repo info} presents the results obtained using ``No Global-Aware'' and four different top-k values for retrieving global functions.
The \textit{No Global-Aware} has both precision and recall as 0 because of lacking of global-aware information as we expected.
Consequently, the generated code only utilizes local functions to accomplish tasks, and it resorts to defining new functions for specific logic if local functions are insufficient.
When k=1, the model shows some awareness of global functions to a certain extent, underlined by the precision and recall as 0.372 and 0.533, respectively.
At k=5, the model's performance significantly improves, indicating that the model now has a more comprehensive understanding of global-aware information.
However, when k=10, the model's F1 score drops by 12.48\%, indicating a substantial reduction in the model's ability to reuse global functions.
As k rises to 15, the effectiveness of the model decreases further.
This is because the overdosed amount (between 10 to 20 functions) of functions is retrieved, introducing excessive interference and hindering accurate reusing.

To further assess the impact of the number of global functions presented to the model, we compute the average number of retrieved functions, denoted as \textit{Avg. Retrieved Functions}.
We find that providing the model with around 8 functions yields the best performance in terms of code reuse, striking a subtle balance.
Thus, we select k=5 as optimal configuration for subsequent experiments.

To further evaluate the effectiveness of our \textit{What-if Code Generator Unit}, we conduct an experiment using the benchmark.
The results showed that the \textit{FunDes} method retrieved 41.5\% of global functions, while the \textit{FunDes+FunCode} method retrieved 60.6\%, representing an almost 20\% increase.
This demonstrates that our \textit{What-if Code Generator Unit} enhances the \textit{Global Function Retrieval Unit} by providing intermediate code for code similarity metrics, thus retrieving more relevant global functions.

\begin{tcolorbox}
\textbf{Finding 2:}
\emph{Global-Aware Information significantly enhances the model's ability to reuse functions in other code files in the repository. The optimal configuration for retrieving global functions is k=5.
The What-if Code Generator Unit helps retrieve more relevant global functions, further enhancing the Global Function Retrieval Unit.}

\end{tcolorbox}

\subsection{RQ3: Does Third-Party Library Aware Knowledge significantly impact LLMs' awareness of available third-party libraries during code generation?}

\subsubsection{Motivation}
In section~\ref{sec:third-party lib knowledge}, we delve into Third-Party-Library Aware information, which helps the model recognize pre-installed third-party libraries in the code repository.
In this section, we aim to evaluate the effectiveness of third-party-library knowledge.

\subsubsection{Experiment Methodology}
To validate this point, we compare code generation between with and without third-party-library aware information. Only local-aware and global-aware information is fed into the model.
Then, we run these two different configurations on the benchmark to obtain code generation results.
Notably, we introduce the \textit{Library Coverage}, the percentage of third-party libraries used in the generated code in the third-part-library base, to further elaborate the extent of reusing.

%%----------------------------
%
%

\begin{table}[t]
    \centering
    \caption{Compare the difference between having Third-Party-Library Aware and not having it on Third-Party Library Reuse Awareness and Library Coverage.}

    \label{table:RQ3 third party lib compare}
    \resizebox{0.46\textwidth}{!}{%
    \begin{tabular}{|c|c|c|c|c|c|}
        \hline \makecell[c]{3rd-Party-Lib\\Awareness}  & Precision & Recall & F1 & Accuracy & \makecell[c]{Library\\Coverage}\\ \hline
        No & 0.716 & 0.429 & 0.536 & 0.715 & 0.875 \\ \hline       
        Yes & \textbf{0.935} & \textbf{0.585} & \textbf{0.720} & \textbf{0.825} & \textbf{0.940} \\ \hline

    \end{tabular}
    }%

\end{table}

%
%
%%----------------------------

\subsubsection{Result Analysis}

Table~\ref{table:RQ3 third party lib compare} demonstrates that after incorporating Third-Party-Library Aware Knowledge, the model exhibits significant improvements in precision, recall, F1 score, and accuracy, with increases of 30.59\%, 36.36\%, 34.33\%, and 15.38\%, respectively.
This enhancement can be attributed to the fact that, with Third-Party-Library Aware information, the model is suggested to use a specific subset of libraries, narrowing down its choice range and making it more likely to generate desired code.
Our finding is further proven by that methods with Third-Party-Library Aware information outperform those without such knowledge in the Library Coverage metric by 7.43\%. 
However, library coverage does not reach 100\% due to occasional instances where the model obstinately believes that there are no viable libraries in the third-party repository to implement the code, which could be attributed to model hallucinations~\cite{ji2023survey,zhang2023language,roller-etal-2021-recipes}.
To mitigate these hallucinations, one potential approach is to use a reflective approach~\cite{Ren2023FromMT, Si2024AST, Shinn2023ReflexionLA}. 
If the model uses a non-pre-installed library, we can inform the model that the library is not expected and prompt it to regenerate the code.
This process is repeated until the model uses a pre-installed library.
Although we have not systematically validated this approach due to the high labeling effort and API token costs, it represents a feasible method to mitigate model hallucinations.

\begin{tcolorbox}

\textbf{Finding 3:} \emph{Third-Party-Library Aware information significantly improves the model's ability to utilize available third-party libraries, reducing risk of potential compatibility issue.}

\end{tcolorbox}

\subsection{RQ4: What is the overall performance of our \texorpdfstring{$A^3$}-CodGen framework?}
\label{sec:RQ4 result}

\subsubsection{Motivation}
In previous sections, we have individually verified the effectiveness of local-aware, global-aware, and third-party-library-aware information.
We also explored the optimal configurations.
In this section, we aim to integrally validate the effectiveness of our $A^3$-CodGen framework from various perspectives.
Therefore, we conduct an ablation study to evaluate $A^3$-CodGen in terms of local reuse, global reuse, and third-party reuse.
Additionally, as more powerful models are released, we aim to compare the performance of $A^3$-CodGen across different foundation models.

\subsubsection{Experiment Methodology}

We first compare our $A^3$-CodGen with three baselines using the GPT-3.5-Turbo-16k model~\cite{GPT-3.5-Turbo-16k}: 
vanilla ChatGPT, $A^1$ Generation (Local-Aware), and $A^2$ Generation (Local-Aware and Global-Aware).
Then, to investigate whether performance improvements are significant with enhanced model capabilities, we use the latest LLM model, GPT-4o~\cite{GPT4o}, as the foundation model for $A^3$-CodGen and compare its performance against GPT-3.5-Turbo-16k.
We evaluate our framework using four metrics:
reuse awareness, reuse correctness, library coverage, and lines of code (LOC).
The F1 score serves as a comprehensive measure of reuse awareness. 
Notably, a lower LOC signifies more concise code.

\begin{table*}[t]
    \centering
    \caption{Comparison of the overall effect of the framework with ablation baselines using GPT-3.5-Turbo-16k for $A^1$-CodGen, $A^2$-CodGen, and $A^3$-CodGen configurations, and comparing $A^3$-CodGen with GPT-3.5-Turbo-16k and GPT-4o.}

    \label{table:RQ4 aaa generation}
    \resizebox{0.9\textwidth}{!}{%
\begin{tabular}{|cc|c|c|c|c|c|}
\hline
\multicolumn{2}{|c|}{Metrics} & \begin{tabular}[c]{@{}c@{}}Vanilla\\GPT-3.5-Turbo-16k\end{tabular} & \makecell[c]{$A^1$-CodGen\\(GPT-3.5-Turbo-16k)} & \makecell[c]{$A^2$-CodGen\\(GPT-3.5-Turbo-16k)} & \makecell[c]{$A^3$-CodGen\\(GPT-3.5-Turbo-16k)} & \makecell[c]{$A^3$-CodGen\\(GPT-4o)} \\ \hline
\multicolumn{1}{|c|}{\multirow{2}{*}{\makecell[c]{Local\\Reuse}}} & \makecell[c]{Awareness (F1)} & N/A & 0.673 & 0.673 & 0.694 & \textbf{0.718} \\ \cline{2-7} 
\multicolumn{1}{|c|}{} & \makecell[c]{Correctness (F1)} & N/A & 0.675 & 0.645 & 0.632 & \textbf{0.879} \\ \hline

\multicolumn{1}{|c|}{\multirow{2}{*}{\makecell[c]{Global\\Reuse}}} & \makecell[c]{Awareness (F1)} & N/A & N/A & 0.601 & 0.606 & \textbf{0.701}  \\ \cline{2-7} 
\multicolumn{1}{|c|}{} & \makecell[c]{Correctness (F1)} & N/A & N/A & 0.611 & 0.591 & \textbf{0.676} \\ \hline

\multicolumn{1}{|c|}{\multirow{3}{*}{\begin{tabular}[c]{@{}c@{}}3rd-Party\\Library\\Reuse\end{tabular}}} & \makecell[c]{Awareness (F1)} & 0.635 & 0.652 & 0.536 & 0.720 & \textbf{0.898} \\ \cline{2-7} 
\multicolumn{1}{|c|}{} & \makecell[c]{Correctness (F1)} & 0.644 & 0.734 & 0.735 & 0.852 & \textbf{0.894} \\ \cline{2-7} 
\multicolumn{1}{|c|}{} & \makecell[c]{Library Coverage} & 0.786 & 0.863 & 0.875 & 0.940 & \textbf{0.976} \\ \hline

\multicolumn{2}{|c|}{\makecell[c]{Avg Line of Code (LOC)}} & 16.308 & 16.507 & \textbf{12.598} & 12.804 & 13.885 \\ \hline
% \multicolumn{2}{|c|}{CodeBLEU} & 0.339 & \textbf{0.344} & 0.312 & 0.322 \\ \hline
\end{tabular}
    }

\end{table*}
%
%
%%----------------------------

\subsubsection{Result Analysis}

Upon evaluating the awareness aspect, as shown in Table~\ref{table:RQ4 aaa generation}, $A^3$-CodGen based on GPT-3.5-Turbo-16k demonstrates excellence in local reuse, global reuse, and third-party library reuse.
This underscores the effective assistance provided by the three categories of repository-aware knowledge in enabling the model to be cognizant of repository information during code generation.

In terms of correctness, we observe that Local Reuse Correctness is optimal for $A^1$ Generation (Local-Aware), Global Reuse Correctness is best for $A^2$ Generation (Local-Aware and Global-Aware), and Third-Party Library Reuse Correctness is highest for $A^3$-CodGen.
This suggests that increasing input knowledge may influence the model's performance in specific reuse scenarios.
Nevertheless, $A^3$-CodGen exhibits only a slight difference from the baseline in local reuse and global reuse correctness (0.003-0.015). 
Despite this minor difference in correctness metrics and its superior performance in other aspects, $A^3$-CodGen demonstrates robust overall performance in local reuse, global reuse, and third-party library reuse.

Looking at the Library Coverage, $A^3$-CodGen achieves near-perfect performance with a score of 0.940, signifying its ability to effectively reuse pre-installed libraries and avoid compatibility issues such as “ModuleNotFoundError: No module named xxx.”
Furthermore, the framework consistently generates code with an optimal number of lines (LOC), indicating its capability to produce concise and efficient code.

By examining the last two columns (Table~\ref{table:RQ4 aaa generation}) of $A^3$-CodGen with GPT-3.5-Turbo-16k and GPT-4o, we observe significant improvements in local reuse, global reuse, and third-party library reuse for GPT-4o compared to GPT-3.5-Turbo-16k. 
This demonstrates that the same prompts achieve better performance as the foundation model's capabilities improve. 
Thus, our $A^3$-CodGen's performance enhances with the increased ability of the foundation model.
More powerful foundation models in future will lead to even better results.
However, the code generated by $A^3$-CodGen based on GPT-4o is slightly more complex, with an average LOC increase of one line, compared to GPT-3.5-Turbo-16k.
This is because GPT-4o considers more logical coherence between code segments, resulting in slightly more complex code. 
Nonetheless, the generated code remains more concise than that produced by Vanilla GPT-based code generation.

\begin{tcolorbox}

\textbf{Finding 4:} \emph{
$A^3$-CodeGen framework excels in awaring three types of code repository knowledge, resulting in an excellent performance in terms of local reuse, global reuse, and third-party library reuse.
Thanks to its outstanding reuse capabilities, our framework can generate concise code that aligns with the code repository's working environment.
Additionally, $A^3$-CodGen's performance improves with the increased capabilities of the foundation model.
}
\end{tcolorbox}

\subsection{RQ5: How does \texorpdfstring{$A^3$}-CodGen framework compare to GitHub Copilot?}

\subsubsection{Motivation}
Given advancements in AI-driven code generation, it is essential to compare our $A^3$-CodGen framework with leading tools like GitHub Copilot~\cite{GithubCopilot}.
GitHub Copilot generates code by leveraging local and global contextual information from neighboring tabs,\footnote{\href{https://github.blog/ai-and-ml/github-copilot/how-github-copilot-is-getting-better-at-understanding-your-code/}{https://github.blog/ai-and-ml/github-copilot/how-github-copilot-is-getting-better-at-understanding-your-code/}} similar to our context-aware mechanisms.
If Copilot can outperform our approach, further development of our method may be unnecessary. However, if it cannot, our goal is to identify the limitations in Copilot's design for collecting context and explore how to enhance its code reuse capabilities using insights from our framework.

\subsubsection{Experiment Methodology}
For each function in the benchmark dataset, we locate the function within the repository using the repository name and fully qualified name, then remove the function code and the corresponding import statements.
We provide the function's comment and definition to trigger GitHub Copilot to generate the code by pressing the `tab' key.
To assist Copilot in generating third-party library imports, we add a comment \textit{\#generate import statements if necessary:}. 
If Copilot makes no suggestion, it indicates no import statements are needed. 
We retain all necessary local functions that could be reused within the function. 
Finally, we used the same labeling method mentioned in Section~\ref{sec_metrics} to label the generated code.
The GitHub Copilot plugin version in this experiment is 1.5.15.6094, and we disable the auto-update in settings.
We compare $A^3$-CodGen with GitHub CoPilot in terms of local reuse, global reuse, and third-party library reuse.
We use the \textit{RepoEval} dataset mentioned in Section~\ref{sec_benchmark_construction}.

\begin{figure}%[H]
    \centering
    \includegraphics[width=0.48\textwidth]{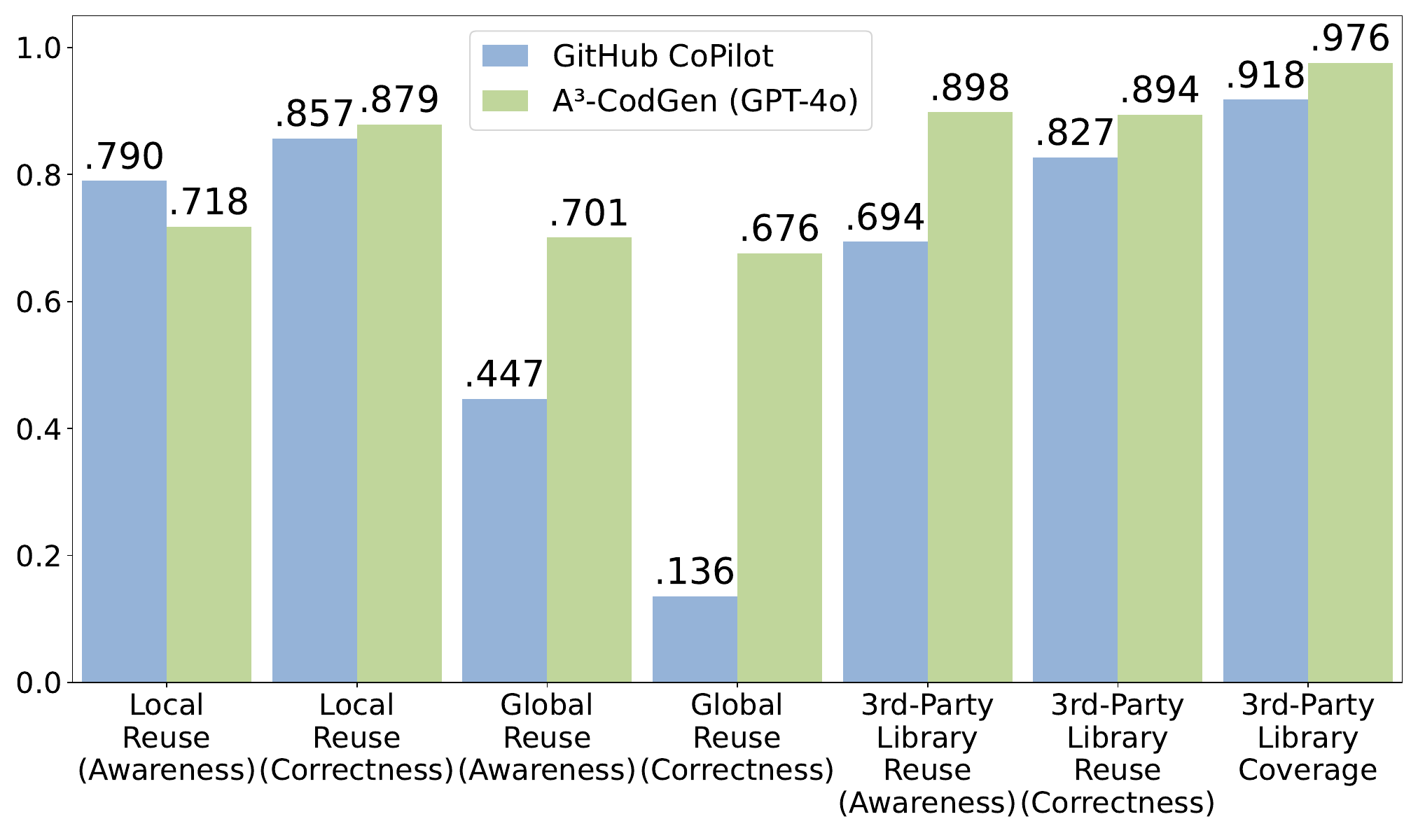}
    \caption{Comparison between GitHub CoPilot and $A^3$-CodGen.}
    \label{fig_gpt4_vs_copilot}

\end{figure}

\subsubsection{Result Analysis}
To evaluate the statistical significance of the differences between $A^3$-CodGen and CoPilot, we conduct Welch’s t-test~\cite{Bl1947THEGO} with a significance threshold of $p < 0.05$.
The p-values for local reuse awareness and correctness are 0.0696 and 0.5590, indicating no significant difference. 
For global reuse awareness and correctness, the p-values are 0.0144 and 1.46e-06, and for third-party library reuse awareness, correctness, and coverage, the p-values are 1.39e-11, 0.0231, and 0.0176, indicating significant differences.

Fig.~\ref{fig_gpt4_vs_copilot} compares GitHub Copilot and $A^3$-CodGen. 
For local reuse, GitHub CoPilot performs excellently with an F1 score of 0.790 for local reuse awareness, while $A^3$-CodGen achieves an F1 score of 0.718. 
This is because CoPilot reviews all working context details (including function definitions, function code, and descriptions) when generating the code. 
In terms of local reuse correctness, $A^3$-CodGen reaches an F1 score of 0.879, slightly better than GitHub CoPilot's 0.857, due to our structured approach to local knowledge.

However, for global reuse awareness and correctness, CoPilot's performance is suboptimal, with F1 scores of 0.447 and 0.136, respectively. 
CoPilot's workflow, although capable of utilizing neighboring tabs' context, struggles with global function calls as necessary global functions might exist in unopened tabs.
Opening all repository tabs is impractical, and CoPilot does not structure information for completion, leading to chaotic context.
In contrast, $A^3$-CodGen retrieves relevant global functions and structures them, facilitating easier and more accurate reuse.
For third-party library reuse, $A^3$-CodGen outperforms CoPilot in awareness, correctness, and coverage.
$A^3$-CodGen helps the model to be aware of all pre-installed libraries, whereas GitHub CoPilot randomly uses third-party libraries.
While CoPilot excels in local information reuse, it lacks in global functions and third-party libraries reuse.

Our method has potential to enhance Copilot by using knowledge retrieval to fetch relevant global functions and library information from the repository, enabling repository-aware code generation. 
For instance, Copilot's limitations can be mitigated by incorporating interpreter information from the compiler to help Copilot become aware of third-party libraries. 
Additionally, using global function retrieval methods can assist Copilot in reusing functions from other code files, not just from neighboring tabs.
Structuring the information before passing it to Copilot, rather than providing all information together, may also improve performance.

\begin{tcolorbox}
\textbf{Finding 5:} \emph{
GitHub CoPilot excels in local information reuse, but its global and third-party library reuse is suboptimal. 
$A^3$-CodGen performs comparably to CoPilot in local reuse while significantly outperforming in global and third-party library reuse due to structured knowledge retrieval. 
Our $A^3$-CodGen can potentially enhance AI-assistant tools like GitHub CoPilot for better code generation.
}
\end{tcolorbox}

\section{Discussion}

\subsection{New Paradigm for Code Generation Tasks}
Code generation tasks have traditionally focused on automating code production for specific requirements, with advancements like GitHub CoPilot and ChatGPT mastering this task.
However, real-world code generation extends beyond meeting requirements;
it requires generating code aligned with project-specific needs and leveraging existing knowledge for improved readability, maintainability, and practicability. 
Therefore, we hope this work sheds a light on this emerging but significant code generation topic:
generating context-aware code that is compatible with the working environment.

\subsection{Knowledge Engineering for Code Generation}
LLMs are sensitive to prompts and well-designed prompts are decisive in obtaining high-quality results~\cite{Yang2021RecentAI,Shin2021ASO,Hayati2018RetrievalBasedNC}.
Either unorganized or too lengthy prompts are impractical.
Instead, it is advisable to extract semantic-enriched knowledge from the code repository to guide the model to directional generate code. 
However, how to selectively fuse the different types of knowledge presents a challenge, potentially impacting the model's awareness of different knowledge types.
In Section~\ref{sec:RQ1 result}, we find that providing diverse information does not necessarily improve model performance.
Meanwhile, in section~\ref{sec:RQ4 result}, we find that providing the model with global awareness and third-party library information unexpectedly enhances its ability to reuse local functions.
This underscores the interaction of diverse code knowledge types and stresses the significance of efficiently mining and utilizing code knowledge in future studies.

\subsection{Dataset Construction for Context Aware Code Generation}
\subsubsection{Benchmark Representativeness}
Experiments use 29 meticulously selected Python code repositories, emphasizing high quality indicated by stars and branches, while avoiding data leaks.
These repositories collectively represent the Python programming ecosystem.
Labeling all functions is highly labor-intensive. Therefore, as described in Section~\ref{sec_benchmark_construction}, we perform sampling to ensure the estimated accuracy within the sample remains within a 0.05 error range while including functions from each code repository to ensure representativeness, with an average of 13.2 functions per repository.

\subsubsection{Benchmark Construction for Different Contexts}
In this paper, we validate the model's awareness of local functions, global functions, and third-party libraries. Future studies could include a more diverse range of repositories to test the framework's effectiveness in different contexts, such as awareness of classes, inherited functions, and variables. The dataset construction method can follow the approach in Section~\ref{sec_benchmark_construction}, where we first label functions to create a benchmark, then evaluate the generated code by predicting labels, and finally calculate reuse awareness and correctness.

In the future, we will maintain and update this dataset, providing the community with a valuable resource for evaluating repository-aware code generation.

\subsection{Threats to Validity}
The threats to validity of this paper lies in the following aspects.

\subsubsection{Subjective Labeling}
Our study involves manual labeling, introducing subjectivity in annotations.
We mitigate this by engaging two professional developers for labeling, resolving any instances with ambiguous labels to establish a reliable ground truth. 
The dataset is public for community validation~\cite{our_repo}.

\subsubsection{Library Dependencies Control}
Due to the complexity of library dependencies, constructing all 29 repository library dependencies is challenging.
Thus, we do not consider the library version for $A^3$ code generation in our experiments.
However, users can feasibly extract third-party libraries and generate code using the `pip list' command or by checking the Python Interpreter in their environment to specify specific versions of libraries, as mentioned in Section~\ref{sec:third-party lib extraction}.

\subsubsection{Generalizing the A3-CodGen Framework to Other Programming Languages}
While the experiments in this paper are conducted on Python code repositories, our $A^3$-CodGen framework is not limited to Python.
To adapt it to another language, such as Java, developers need to use language-specific program analysis tools (e.g., Spoon~\cite{Spoon} for Java) for the \textit{Repository Knowledge Base Construction}.
This involves extracting relevant repository knowledge, such as signatures and source code. 
Other steps, including \textit{repository-aware knowledge retrieval} and \textit{$A^3$ code generation}, remain unchanged.
Although constructing the knowledge base may require additional engineering effort, it does not involve new scientific innovations.
It simply requires following our outlined approach using the appropriate program analysis tools for the specific language.
\section{Related Work}~\label{sec_related}
Code generation tasks is about creating source code from provided natural language descriptions or requirements, which has been a longstanding focal point in SE research.

\textbf{Repository-Level Code Completion.}
Repository-level code completion is gaining traction over traditional function-level completion~\cite{Shrivastava2022RepositoryLevelPG, Shrivastava2023RepoFusionTC, Ding2022CoCoMICCC, Zhang2023RepoCoderRC}.
Shrivastava et al. generated example-specific prompts using a prompt proposal classifier~\cite{Shrivastava2022RepositoryLevelPG} and further proposed \textit{RepoFusion} to incorporate relevant repository context by training code models~\cite{Shrivastava2023RepoFusionTC}.
In addition, 
Ding et al.~\cite{Ding2022CoCoMICCC} proposed a cross-file context finder tool, named \textit{CCFinder}, which constructs a project context graph and retrieves the most relevant cross-file context based on local import statements.
\textit{RepoCoder}~\cite{Zhang2023RepoCoderRC} retrieved similar code snippets from the repository as part of a prompt and proposes an iterative retrieval-generation paradigm.
This novel strategy bridged the gap between the retrieved context and the intended target through an interactive process.

\textbf{Retrieval Augmented Code Generation.}
Retrieval Augmented Code Generation (RACG)~\cite{Drain2021GeneratingCW,Hayati2018RetrievalBasedNC,Parvez2021RetrievalAC,Zhou2022DocPromptingGC} represents an emerging and promising area of research that combines the strengths of retrieval-based methods with code generation techniques.
Hayati et al.\cite{Hayati2018RetrievalBasedNC} introduced \textit{RECODE}, the first model to integrate information retrieval into code generation. 
Parvez et al.~\cite{Parvez2021RetrievalAC} proposed the \textit{REDCODER}, which utilizes retrieval to enhance the input of the generation model, aiming to maximize the utilization of additional information.
Drain et al.~\cite{Drain2021GeneratingCW} retrieve template functions from GitHub and relevant answers from Stack Overflow to improve code generation. 
Zhou et al.~\cite{Zhou2022DocPromptingGC} proposed Docprompting, which retrieve relevant APIs from documents and add them into prompts to improve code generation. 
Zan et al.~\cite{Zan2022WhenLM} proposed CodeGenAPI, which retrieve private APIs from documents to improve code generation. 
The APIRetriever module of CodeGenAPI and our approach both focus on API recommendation.
However, CodeGenAPI directly retrieves APIs from documentation and cannot effectively handle high-level tasks that require combinations of APIs.

\textbf{Code Completion Evaluation.}
LLM-based code completion and program synthesis methods provide unprecedented convenience, however researcher also question the actual performance under the mirage from different perspectives~\cite{yuan2023evaluating, liu2023your, du2023classeval, ouyang2023llm, yu2023codereval}.
Yuan et al. evaluated 10 open-source instructed LLMs on four representative code comprehension and
generation tasks~\cite{yuan2023evaluating}.
Liu et al. proposed \textit{EvalPlus}, a code synthesis benchmarking framework, to rigorously evaluate the functional correctness of LLM-synthesized code in function level~\cite{liu2023your} .
In a more challenging code generation scenario, i.e., class-level code generation, Du et al. manually constructed the first class-level
code generation benchmark \textit{ClassEval} of 100 class-level Python code generation tasks~\cite{du2023classeval}.
In addition, Ouyang et al. conducted an empirical study to demonstrate that non-determinism in code generation tasks~\cite{ouyang2023llm}.
Yu et al. proposed a benchmark named \textit{CoderEval}, which can be used to assess the performance of models against pragmatic code generation beyond just generating standalone functions~\cite{yu2023codereval}.

\section{Conclusion and Future Work}
Code generation within code repositories often necessitates comprehensive knowledge of the entire repository context. Existing techniques often lack such a holistic understanding of code repository knowledge, which is crucial for effective code generation.
In this paper, we introduce the $A^3$-CodGen framework, which is designed to construct, mine, and unify three essential types of code repository knowledge: 
local-aware, global-aware, and third-party-library-aware knowledge.
To validate the effectiveness of these knowledge types, we conducted benchmark experiments and identified optimal configurations for their utilization.
The results demonstrate that the $A^3$-CodGen framework significantly enhances the ability of LLMs to fully utilize local, global, and third-party library information. This results in the generation of concise, high-quality code.
In the future, we plan to expand the applicability of our framework to multiple programming languages.
Furthermore, we aim to enhance our knowledge utilization methods to enable LLMs to become more aware of various types of code knowledge within code repositories.

\normalem
\balance
\bibliography{samplebase}

\end{document}